\documentclass[twocolumn,showpacs,prc,floatfix]{revtex4}
\usepackage{graphicx}
\usepackage{amsmath}
\newcommand{\p}[1]{(\ref{#1})}
\newcommand\beq{\begin{eqnarray}} \newcommand\eeq{\end{eqnarray}}
\newcommand\beqstar{\begin{eqnarray*}} \newcommand\eeqstar{\end{eqnarray*}}
\newcommand{\beqe}{\begin{equation}} \newcommand{\eeqe}{\end{equation}}
\newcommand{\bal}{\begin{align}}
\newcommand{\fr}[2]{\frac{\partial{#1}}{\partial{#2}}}
\newcommand{\tbf}{\textbf}
\newcommand{\tit}{\textit}

\begin {document}
\title{Anisotropic multi--gap superfluid states \protect\\  in
nuclear matter }
\author{ A. A. Isayev}
\affiliation{Kharkov Institute of Physics and Technology,
Academicheskaya Str. 1,
 Kharkov, 61108, Ukraine}
 \author{G. R\"opke}
 \affiliation{University of Rostock, FB Physic, Universit\"atplatz 3,  Rostock, 18051,
 Germany}
 \date{\today}
\begin{abstract}It is
 shown that under changing  density or temperature a nucleon
Fermi superfluid can undergo a phase transition to an anisotropic
superfluid state, characterized by nonvanishing gaps in pairing
channels with  singlet--singlet (SS) and triplet-singlet (TS)
pairing of nucleons (in spin and isospin spaces). In the SS pairing
channel   nucleons are paired with nonzero orbital angular
momentum. Such two--gap states can arise as a result of branching
from the one-gap solution of the self-consistent equations, describing
SS or TS pairing of nucleons, that depends on the relationship
between SS and TS coupling constants at the branching point. The
density/temperature dependence of the order parameters  and the
critical temperature for transition to the anisotropic two--gap state
are determined in a model with the SkP effective interaction. It is
shown that the anisotropic SS--TS superfluid phase corresponds to
a metastable state in nuclear matter.
\end{abstract}
 \maketitle

\section{Introduction}
It is known that at sufficiently low temperatures a nucleon Fermi
system becomes unstable with respect to formation of Cooper pairs
due to the attractive component of the nucleon--nucleon (NN) potential.
Basically in early articles on superfluidity
isovector pairing of the same nucleons in nuclei and nuclear
matter~\cite{BM,RS} was studied. Later it was realized that isoscalar
neutron--proton pairing plays an essential role in the description
of superfluidity of finite nuclei with $N\simeq Z$ (see
Refs.~\cite{G,RSSL} and
 references therein) and nearly symmetric nuclear matter
 \cite{AFRS}--\cite{SAL}. Possible coexistence of isoscalar and
 isovector nucleon pairing in finite nuclei was studied in
 Ref.~\cite{G1}. In addition to isovector pairing, isoscalar
 pairing modes have been investigated to explain the excitation
 energies in $N=Z$ nuclei~\cite{SW}.
 As was shown in Ref.~\cite{AIP2}, at low densities of nuclear matter
the coupling between isospin $T=0$ and $T=1$ pairing channels may
be of importance, leading to the emergence of multi--gap
superfluid states  with nonvanishing gaps in both pairing
channels.  In Refs.~\cite{AIP2,AAI} the case of a multi--gap
condensate was considered, when the energy gaps in $^1S_0$ and
$^3S_1$ pairing channels are nonzero. For such a condensate the
orbital angular momentum $L$ of a pair is equal to zero and,
hence, the order parameters in both pairing channels are isotropic
functions of the momentum. However, we can consider another
possibility, when the pairing of nucleons in one (or both) pairing
channels occurs in a state with nonzero orbital angular momentum.
In this case an anisotropic multi--gap condensate will be
realized, where the order parameter(s) will depend not only on the
absolute value of momentum, but also on its direction. Further we
will consider an isoscalar multi--gap superfluid state,
corresponding to a superposition of pair states with spin $S=0$
and $S=1$. According to the Pauli principle, the orbital angular
momentum in the first pairing channel ($S=0,T=0$) is odd and in
the second one ($S=1,T=0$) it is even. Hence, such a multi--gap
condensate will necessarily be anisotropic. Our main purpose will
be the clarification of a mechanism for the appearance of
anisotropic multi--gap superfluid states, finding the
corresponding critical temperature and comparing the free energies
for different superfluid phases. As a theoretical framework, we
use the Fermi--liquid (FL) approach ~\cite{AKP}, in which normal
and anomalous FL interaction amplitudes are treated on an equal
footing. As a NN interaction we choose the density dependent
Skyrme effective forces, used earlier in a number of contexts for
calculations in finite nuclei~\cite{DFT,RDN} and infinite nuclear
matter~\cite{SYK}--\cite{AIPY}. In specific calculations we use
the SkP potential~\cite{DFT}, for which the strongest interaction
in the state of a Cooper pair with nonzero orbital angular
momentum is realized in  the pairing channel $S=0,T=0$, and in the
state with $L=0$ it is in the channel  $S=1,T=0$.
\section{Basic Equations}
 Superfluid states of nuclear matter are described
  by the normal $f_{\kappa_1\kappa_2}=\mbox{Tr}\,\varrho
  a^+_{\kappa_2}a_{\kappa_1}$ and
 anomalous $g_{\kappa_1\kappa_2}=\mbox{Tr}\,\varrho
a_{\kappa_2}a_{\kappa_1}$ distribution functions of nucleons
($\kappa\equiv({\bf{p}},\sigma,\tau)$,  ${\bf p}$ is momentum,
$\sigma(\tau)$ is the projection of spin (isospin) on the third
axis, $\varrho$ is the density matrix of the system).  The energy
of the system is specified as a functional of the distribution
functions $f$ and $g$, $E=E(f,g)$. It determines the quasiparticle
energy $\varepsilon$ and the matrix order parameter $\Delta$ of
the system
 \beq
\varepsilon_{\kappa_1\kappa_2}=\fr{E}{f_{\kappa_2\kappa_1}},\quad
\Delta_{\kappa_1\kappa_2}=2\fr{E}{g_{\kappa_2\kappa_1}^+}\,\,.
\label{1}
\eeq
The self-consistent matrix equation for determining
the distribution functions $f$ and $g$ follows from the minimum
condition of the thermodynamic potential \cite{AKP} and is
  \beq
\hat f=\left\{\mbox{exp}(Y_0\hat\varepsilon+\hat
Y_4)+1\right\}^{-1}\equiv
\left\{\mbox{exp}(Y_0\hat\xi)+1\right\}^{-1},\label{2}\eeq
$$
\hat f=\left(\begin{array}{cc}f&g\\g^+&1-f^{\mbox{\small T}}
\end{array}\right),\;
\hat\varepsilon=\left(\begin{array}{cc}\varepsilon&\Delta\\ \Delta^+&
-\varepsilon^{\mbox{\small T}}\end{array}\right),\; $$ $$\hat
Y_4=\left(\begin{array}{cc}Y_4&0\\
0& -Y_4\end{array}\right)\,\,.
$$
Here  the quantities $\varepsilon,\Delta,Y_4$ are, in turn,
matrices in the space of the $\kappa$ variables, with
$Y_{4\kappa_1\kappa_2}=Y_{4\tau_1}\delta_{\kappa_1\kappa_2}$
$(\tau_1=p,n)$, $Y_0=1/T,\ Y_{4p}=-\mu_p/T$ and $Y_{4n}=-\mu_n/T$
are the Lagrange multipliers, $\mu_p$ and $\mu_n$ are the chemical
potentials for protons and neutrons, $T$ is the temperature. We shall
study two--gap superfluid states in symmetric nuclear matter,
corresponding to a superposition of pair states with total spin $S$ and
isospin $T$: $S=0$, $T=0$ (singlet--singlet (SS) pairing
in spin and isospin spaces) and $S=1,T=0$ (triplet--singlet (TS)
pairing) with the projections $S_z=T_z=0$ (SS--TS states).
 In this case the
normal $f$ and anomalous $g$ distribution functions
read~\cite{AIP} \bal f({\bf p})&= f_{00}({\bf p})\sigma_0\tau_0
,\label{7.2}\\
g({\bf p})&=g_{00}({\bf p})\sigma_2\tau_2+g_{30} ({\bf
p})\sigma_3\sigma_2\tau_2,\nonumber
\end{align}
  where $\sigma_i$ and $\tau_k$ are the Pauli
matrices in spin and isospin spaces, respectively. The components
of the anomalous distribution function in Eq.~\p{7.2} possess
different symmetry properties,
\beqe 
 g_{00}(-{\bf p})=-g_{00}({\bf p}),\;
g_{30}(-{\bf p})=g_{30}({\bf p}),
 \end{equation}
 and, hence, considering this, the multi--gap condensate will be anisotropic.
 For the energy functional, which is invariant with
respect to rotations in spin and isospin spaces, the structure of
the single particle energy and the order parameter is  similar to
that of the distribution functions $f,g$:
\begin{flalign}
\varepsilon({\bf p})&= \varepsilon_{00}({\bf
p})\sigma_0\tau_0,\label{8.2}\\
\Delta({\bf p})&=\Delta_{00}({\bf
p})\sigma_2\tau_2+\Delta_{30}({\bf p})\sigma_3\sigma_2\tau_2,
\nonumber\end{flalign} where
\beqe 
 \Delta_{00}(-{\bf p})=-\Delta_{00}({\bf p}),\;
\Delta_{30}(-{\bf p})=\Delta_{30}({\bf p}).
 \end{equation}

Using the procedure of block diagonalization
 \cite{AKP}, one
 can evidently express the distribution functions $f_{00},g_{00},g_{30}$
 in
terms of the quantities $\varepsilon$ and $\Delta$: \bal
f_{00}&=\frac{1}{2}\left[1 -\frac{\xi}{2E_+}(1-2n_+)-
\frac{\xi}{2E_-}(1-2n_-)\right],
 \label{13} \\
g_{00}&=-\frac{\Delta_+}{4E_+}(1-2n_+)-
\frac{\Delta_-}{4E_-}(1-2n_-),\label{11}\\
g_{30}&=-\frac{\Delta_+}{4E_+}(1-2n_+)+
\frac{\Delta_-}{4E_-}(1-2n_-).\label{12} \end{align} Here
\begin{gather*}
E_\pm=\sqrt{\xi^2+|\Delta_\pm|^2},\quad\Delta_\pm=\Delta_{00}\pm\Delta_{30},\\
\xi({\bf p})=\varepsilon_{00}({\bf p})-\mu_0,\;n_{\pm}=\{\exp
(Y_0E_\pm)+1\}^{-1},
\end{gather*}
$\mu_0$ is the chemical potential,
which should be determined from the normalization condition \beq
\frac{4}{\cal V}\sum_{\bf p}f_{00}({\bf
p})=\varrho,\label{13.1}\eeq  $\varrho$  is density of symmetric
nuclear matter. As follows from Eqs.~\p{13}--\p{12}, the
nucleon superfluid is characterized by two types of fermion
excitations with gaps $\Delta_\pm$ in the
 spectrum.
 In this case the spectrum is two--fold split due to coupling of SS and TS pairing
 channels ($\Delta_{00}\not=0,\Delta_{30}\not=0$).

To obtain the self--consistent equations for the quantities
$\Delta$ and $\xi$, it is necessary to specify the energy
functional of the system, which we write in the form
\begin{equation} E(f,g)=E_0(f)+E_{int}(f)+E_{int}(g), \label{14}
\end{equation}
\bal {E}_0(f)&=4\sum\limits_{ \bf p}^{} \varepsilon_0({\bf
p})f_{00}({\bf p}), \;{E}_{int}(f)=2\sum\limits_{ \bf p}^{}
\tilde\varepsilon_{00}({\bf p})f_{00}({\bf p}), \nonumber\\
\varepsilon_0({\bf p})&=\frac{{\bf p}^{\,2}}{2m_{0}},  \,
\tilde\varepsilon_{00}({\bf p})=\frac{1}{2\cal V}\sum_{\bf
q}U_0({\bf k})f_{00}({\bf q}),\,{\bf k}=\frac{{\bf p}-{\bf q}}{2},
\nonumber\end{align} \beqstar {E}_{int}(g)&=&\frac{2}{\cal V}
\sum\limits_{ {\bf p},{\bf q}}^{} \Bigl(g_{00}^*( {\bf p})V_0(
{\bf p}, {\bf q})g_{00}({\bf q})
\\ &+& g_{30}^*( {\bf p})V_1(
 {\bf p}, {\bf q})g_{30}({\bf q})\Bigr).
\end{eqnarray*}
 Here
  $m_0$ is the bare mass of a nucleon, $U_0({\bf k}) $ is the normal FL
amplitude, $V_0({\bf p},{\bf q}), V_1({\bf p},{\bf q})$ are the
anomalous FL amplitudes, describing interactions in the SS and TS
pairing channels, respectively. With allowance for Eqs.~\p{1},
\p{14}, we obtain self--consistent equations in the form
  \begin{align} \xi({\bf
p})&=\varepsilon_0({\bf p})-\mu_0+\tilde\varepsilon_{00}({\bf
p}),\;
\label{15}\\
 \Delta_{00}({\bf p})&=\frac{1}{\cal V}\sum_{\bf q}V_0({\bf p},
 {\bf
q})g_{00}({\bf q}),\;\label{16}\\
\Delta_{30}(\bf p)&=\frac{1}{\cal V}\sum_{\bf q}V_1({\bf p},{\bf
q})g_{30}({\bf q}).\label{17}
\end{align}
Taking into account
Eqs.~\p{11}, \p{12}, we obtain equations for the energy gaps
$\Delta_{00},\Delta_{30}$
\begin{align}
&\Delta _{00}({\bf
p})=-\frac {1}{ 4\cal{V}}\sum\limits_{\bf q}^{} V_0({\bf p},{\bf q
})\label{12.1}\\ &\times\left\{ \frac{\Delta_+({\bf q})}{E_+({\bf
q)}}\tanh\frac{E_+({\bf q})}{2T}+ \frac{\Delta_-({\bf
q})}{E_-({\bf
q)}}\tanh\frac{E_-({\bf q})}{2T}\right\},\nonumber \\ 
&\Delta _{30}({\bf p})=-\frac {1}{ 4\cal{V}}\sum\limits_{\bf q}^{}
V_1({\bf p},{\bf q })\label{12.2}\\&\times \left\{
\frac{\Delta_+({\bf q})}{E_+({\bf q)}}\tanh\frac{E_+({\bf
q})}{2T}- \frac{\Delta_-({\bf q})}{E_-({\bf
q)}}\tanh\frac{E_-({\bf q})}{2T}\right\}.\nonumber
\end{align}
Eqs.~\p{15}, \p{12.1}, \p{12.2} describe two--gap superfluid
states of symmetric nuclear matter and contain one--gap solutions
with $\Delta_{00}\not=0,\Delta_{30}\equiv0$ (SS pairing) and
$\Delta_{00}\equiv0,\Delta_{30}\not=0$ (TS pairing) as some
particular cases.

To obtain numerical results we will use the Skyrme effective
interaction,  for which the normal and anomalous FL
 amplitudes read~\cite{AIP}
\bal
U_0({\bf k})&=6t_0+t_3\varrho^\beta\\
&\quad+\frac{2}{\hbar^2}[3t_1+t_2(5+4x_2)]{\bf k}^{2},
\nonumber\\
V_0({\bf p},{\bf q})&=\frac{t_2}{\hbar^2}(1-x_2){\bf p}{\bf
q}\equiv V_0(p,q){\bf p}^0{\bf q}^{0},\label{17.1}\\
 V_{1}({\bf p},{\bf q})&=t_0(1+ x_0)
+\frac{1}{6}t_3\varrho^\beta(1+ x_3)\label{17.2}\\
&\quad+\frac{1}{2\hbar^2}t_1(1+ x_1)({\bf p}^2+{\bf q}^{2}),
\nonumber
\end{align}
 where $t_i,x_i,\beta$ are
phenomenological constants, characterizing the given
parametrization of Skyrme forces.
 (We shall use the  SkP
\cite{DFT} potential.) According to Eqs.~\p{17.1}, \p{17.2},
pairing in the SS channel occurs with orbital angular momentum
$L=1$ and in the TS channel with $L=0$. Note that in the case of
the effective Skyrme interaction the normal FL amplitude $U_0$ is
quadratic in momentum and hence describes a renormalization of
free nucleon mass and chemical potential. The expression for the
quantity $\xi$, given by Eq.~\p{15}, with regard for the explicit
form of the amplitude $U_0$, reads
$$\xi=\frac{p^2}{2m}-\mu,$$
where the effective nucleon mass $m$ is defined by the formula
$$\frac{\hbar^2}{2m}=\frac{\hbar^2}{2m_{0}}+\frac{\varrho}{16}
[3t_1+t_2(5+4x_2)]$$ and the effective chemical potential $\mu$
should be determined from  Eq.~\p{13.1}.

\section{Order parameters at zero temperature}
Let us consider the case of zero temperature.  We shall analyze
Eqs.~\p{12.1},\p{12.2}, using the simplifying assumption, that FL
amplitudes $V_0,V_1$ are nonzero
  only in a narrow layer near the Fermi surface:
$|\xi|\le\theta,\, \theta\ll\varepsilon_F$ (we set
$\theta=0.1\varepsilon_F$). Then the TS energy gap represents some
constant quantity $\Delta_{30}\equiv\Delta_{30}(p=p_F)$, while the SS
energy gap is angular dependent, $\Delta_{00}=\Delta{\bf
p}^0{\bf n}$, ${\bf n}$ being an arbitrary real unit vector. In addition,
we shall neglect the influence of the finite size of the gaps on
the chemical potential $\mu$ and set
$\mu=\frac{\hbar^2k_{F}^2}{2m}$,
$k_F=\left(\frac{3\pi^2\varrho}{2}\right)^{1/3}$.  As a result of
these assumptions, we arrive at equations for determining the
quantities $\Delta,\Delta_{30}$:
\bal
\Delta&=\frac{g_0}{2}\int_0^\theta\,d\xi\int_0^1\,dx
x\Bigl(\frac{\Delta
x+\Delta_{30}}{E_+}+\frac{\Delta x-\Delta_{30}}{E_-}\Bigr),\label{18}\\
\Delta_{30}&=\frac{g_1}{2}\int_0^\theta\, d\xi\int_0^1\,dx
\Bigl(\frac{\Delta x+\Delta_{30}}{E_+}-\frac{\Delta
x-\Delta_{30}}{E_-}\Bigr),\label{19}
\end{align} where
$$E_\pm=\sqrt{\xi^2+|\Delta x\pm\Delta_{30}|^2},\quad
g_{0,1}=-\nu_FV_{0,1}(p_F,p_F)$$ and
$\nu_F=\frac{mp_F}{2\pi^2\hbar^3}$ is the density of states at the
Fermi surface for a nucleon with the given spin and isospin
projections. Our main goal is to find two--gap solutions with
$\Delta\not=0,\Delta_{30}\not=0$ and to clarify the mechanism of
their appearance. To give some analytical consideration, we will
assume that the conditions $\Delta,\Delta_{30}\ll\theta$ are
fulfilled (logarithmic approximation). As will become apparent,
this approximation works quite well just in the density region
where the two--gap solutions exist. Introducing the ratio of the
energy gaps $\alpha=\Delta_{30}/\Delta$ and performing integration
in Eqs.~\p{18},\p{19} in the logarithmic approximation, we arrive
at equations for the quantities $\alpha$ and $\Delta$: \bal
\frac{1}{g_1}&=\frac{1}{2}+\ln\frac{2\theta}{|\Delta|}-\frac{1}{2}\ln|\alpha^2-1|
-\frac{1}{4}(\alpha+\frac{1}{\alpha})\ln
\left|\frac{\alpha+1}{\alpha-1} \right|
,\nonumber \\
\frac{3}{g_0}&=\frac{1}{3}+\ln\frac{2\theta}{|\Delta|}-\frac{1}{2}\ln|\alpha^2-1|\label{20}\\
&\quad+\frac{\alpha}{4}(\alpha^2-3)\ln
\left|\frac{\alpha+1}{\alpha-1}
\right|-\frac{\alpha^2}{2}\,\,.\nonumber
\end{align}
Excluding $\Delta$ from Eqs. \p{20}, we obtain the equation
\beqe\frac{1}{g_1}-\frac{3}{g_0}=\varphi(\alpha),\;
\varphi(\alpha)\equiv \frac{1}{2}(\alpha^2+\frac{1}{3})-
\frac{(\alpha^2-1)^2
}{4\alpha}\ln\left|\frac{1+\alpha}{1-\alpha}\right| \label{21}
\end{equation}
Since $\varphi_{min}=\varphi(0)=-1/3$ (the point $\alpha=0$ is the
point of removable discontinuity) and
$\varphi_{max}=\varphi(\pm\infty)=1$, then Eq. \p{21} has a
solution for $\alpha$, if the coupling constants $g_0$ and $g_1$
satisfy the inequalities
 \beqe
-\frac{1}{3}<\frac{1}{g_1}-\frac{3}{g_0}<1.\label{22}
\end{equation}
These restrictions have to be fulfilled in the logarithmic
approximation for the existence of the SS--TS mixed state. The sense of
the restrictions \p{22} on SS and TS coupling constants  is that the
quantities $g_1$ and $g_0/3$ must be of the same order of
magnitude.  Clearly, similar restrictions exist in a general
case, when the conditions of the logarithmic approximation are not
fulfilled. Note that the constraints \p{22} are more strict than
the corresponding ones for the existence of a TS--ST mixed state~
\cite{AIP2}\,:$-1<1/g_1-1/g_2<1,\ g_2=-\nu_FV_2(p_F,p_F)$ (the
constants $g_1,g_2$, determined here, are four times as large as
those in Ref.~\cite{AIP2}).

\begin{figure}[tbp]
\includegraphics[height=7.0cm,width=8.6cm,trim=48mm 142mm 57mm 69mm,
draft=false,clip]{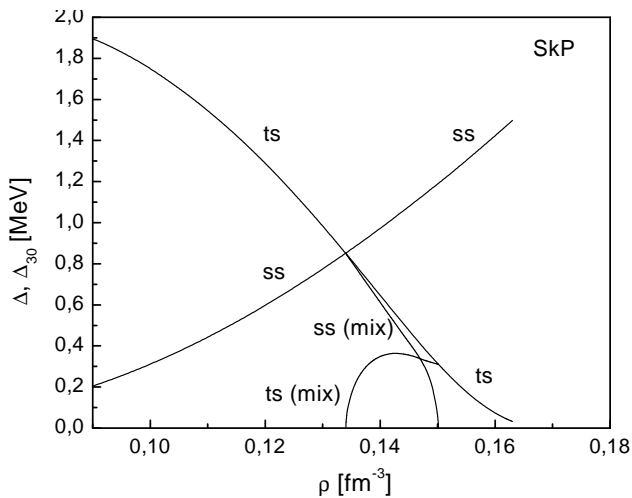} \caption{Order parameters
$\Delta,\Delta_{30}$ vs. density at zero temperature for SkP
force; $ss$(mix) and $ts$(mix) are notations for the dependencies
of the energy gaps $\Delta$ and $\Delta_{30}$ in the mixed SS--TS
solution. }\label{fig1}
\end{figure}

 The results
of a numerical determination of the energy gaps
$\Delta(\varrho),\Delta_{30}(\varrho)$ as functions of density
from Eqs.~\p{18}, \p{19} are presented in  Fig.~\ref{fig1},
where the part of the phase diagram corresponding to the mixed
SS--TS states is shown.
One can see that SS--TS
solutions exist in the finite density region
$\varrho_1<\varrho<\varrho_2$. In the left critical point two--gap
solutions appear  as a result of branching from a one--gap SS
solution (at the branching point
$\Delta_{30}(\varrho_1)=0,\,\,\Delta(\varrho_1)=\Delta^{{ss}}(\varrho_1)
$, $\Delta^{{ss}}$ being a one--gap SS solution). At
$\varrho=\varrho_1$ the coupling constants are related by the
formula
\beqe \frac{1}{g_1(\varrho_1)}-
\frac{3}{g_0(\varrho_1)}=-\frac{1}{3}.\label{23}
\end{equation}
In
the logarithmic approximation the one--gap SS and TS order parameters
have the form
\beqstar \Delta^{ss}=2\theta\exp
\Bigl(-\frac{3}{g_0}+\frac{1}{3}\Bigr),\;
\Delta_{30}^{ts}=2\theta\exp\Bigl(-\frac{1}{g_1}\Bigr).
\eeqstar
 From here and  Eq.~\p{23} it follows that
 at zero temperature at the left branching point
the energy gaps in the SS and TS pairing channels  are equal,
$\Delta^{ss}(\varrho_1)=\Delta^{ts}(\varrho_1)$. This peculiarity
is preserved for finite temperatures, as will be seen in
Section~IV.

 In the right critical point the two--gap solutions
branch off from a one--gap TS solution (at the branching point
$\Delta(\varrho_2)=0,\Delta_{30}(\varrho_2)=\Delta_{30}^{ts}(\varrho_2)$).
At $\varrho=\varrho_2$ \beqe \frac{1}{g_1(\varrho_2)}-
\frac{3}{g_0(\varrho_2)}=1.\label{24}
\end{equation}
Note that the
conditions of the logarithmic approximation,
$\Delta\ll\theta,\Delta_{30}\ll\theta$, are fulfilled quite
satisfactorily in the density domain where the two--gap solutions
exist: the maximum value of the ratios
$\Delta/\theta,\Delta_{30}/\theta$ does not exceed $0.26$.

From Fig.~\ref{fig1} it is seen that SS--TS mixed states exist in
a density interval that is much closer to nuclear matter
saturation density than that for TS--ST multi--gap states
\cite{AIP2}, which exist in the region $\varrho<\varrho'$, where
$\varrho'\approx0.05\div0.06\ \mbox{fm}^{-3}$. Hence, there is no
competition between SS--TS and TS--ST superfluid states, which
exist in quite different density domains.

\section{Critical temperature, Order parameters at nonzero temperature}

The analysis given in the previous section  relates to the case of
zero temperature. It is clear that if SS--TS states exist at T=0,
then such states appear first at some critical temperature. To
determine the critical temperature we use the following
considerations. Obviously, SS--TS solutions arise as a result of
branching from SS or TS one--gap solutions of
Eqs.~\p{12.1},\p{12.2}. If branching occurs from a TS solution
then at the critical point
$\Delta(T_c)=0,\Delta_{30}(T_c)=\Delta_{30}^{ts}(T_c)$.
Considering the limit $\Delta\rightarrow 0$ in
Eqs.~\p{12.1},\p{12.2}, we obtain equations for determining $T_c$
\bal 1&=g_1\int_0^\theta \frac{d\xi}{E}
\tanh\frac{E}{2T_c},\quad E=\sqrt{\xi^2+\Delta_{30}^2},\label{25}\\
 1&=\frac{g_0}{3}\int_0^{\theta} d\xi\,
\Bigl\{\frac{\xi^2}{E^3}\tanh\frac{E}{2T_c}+
\frac{\Delta_{30}^2}{2E^2T_c}\frac{1}{\cosh^2\frac{E}{2T_c}}
\Bigr\}. \label{26} \end{align} The first of these equations
determines the temperature behavior of TS energy gap, the second
one determines the critical temperature $T_c$ at which the mixed
SS--TS solution branches from the  one--gap TS solution. If SS--TS
states appear from a SS solution, then
$\Delta_{30}(T_c)=0,\Delta(T_c)=\Delta^{ss}(T_c)$. Considering the
limit $\Delta_{30}\rightarrow 0$ in Eqs.~\p{12.1},\p{12.2}, we
obtain
\bal 1&=g_0\int_0^\theta d\xi\,\int_0^1
dx\,\frac{x^2}{E}\tanh\frac{E}{2T_c},
\; E=\sqrt{\xi^2+\Delta^2x^2},\label{27}\\
 1&=g_1\int_0^{\theta} d\xi\,\int_0^1dx
\Bigl\{\frac{\xi^2}{E^3}\tanh\frac{E}{2T_c}\nonumber\\
&\quad +\frac{\Delta^2x^2}{2E^2T_c}\frac{1}{\cosh^2\frac{E}{2T_c}}
\Bigr\}\,\,. \label{28}
\end{align}
Here the  first  equation
determines the temperature behavior of the SS energy gap, the second
one determines the critical temperature $T_c$ at which the mixed
SS--TS solution branches from the one--gap SS solution.

\begin{figure}[htbp]
\includegraphics[height=7.0cm,width=8.6cm,trim=48mm 148mm 57mm 66mm,
draft=false,clip]{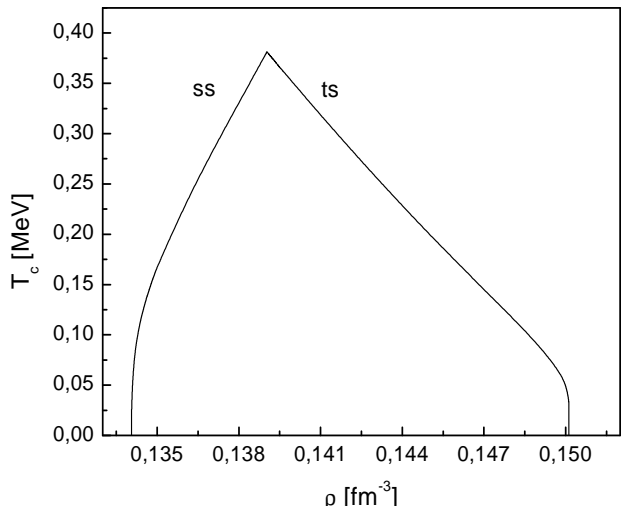} \caption{Critical temperature of
SS--TS superfluid state vs. density for SkP force; notations $ss$
and $ts$ correspond to branching of SS--TS solutions from SS and
TS one--gap solutions, respectively.}\label{fig2}
\end{figure}

The results of a numerical solution of Eqs.~\p{25},\p{26} and
Eqs.~\p{27},\p{28} are presented in Fig.~\ref{fig2}.
One
can see that the curve $T_c(\varrho)$ consists of two branches.
The left branch corresponds to the appearance at a critical temperature
$T_c$  of a SS--TS solution from the SS one--gap solution, the right one
corresponds to the appearance of a SS--TS solution from the TS one--gap
solution. The maximum value of $T_c$ is approximately equal to $0.38$
MeV at density $\varrho_m\approx0.139\;\mbox{fm}^{-3}$. In
the limit $T_c\rightarrow0$, from Eqs.~\p{25},\p{26},
we obtain Eq.~\p{24} for the right critical point
$\varrho=\varrho_2$ ($\varrho_2\approx0.15\; \mbox{fm}^{-3}$) and
from Eqs.~\p{27},\p{28} we obtain Eq.~\p{23} for the left critical
point $\varrho=\varrho_1$ ($\varrho_1\approx0.134\;
\mbox{fm}^{-3}$). Thus, in the density interval
$(\varrho_1,\varrho_m)$ SS--TS solutions appear as a result of a
phase transition in temperature from a one--gap SS solution, and in
the interval $(\varrho_m,\varrho_2)$ they appear from a one--gap TS
solution. If $\varrho_1<\varrho<\varrho_m$, the coupling constants
satisfy the inequality $g_1>g_0/3$, and for
$\varrho_m<\varrho<\varrho_2$ it is $g_1<g_0/3$.

\begin{figure}[htbp]
\includegraphics[height=12.6cm,width=8.6cm,trim=49mm 105mm 53mm 50mm,
draft=false,clip]{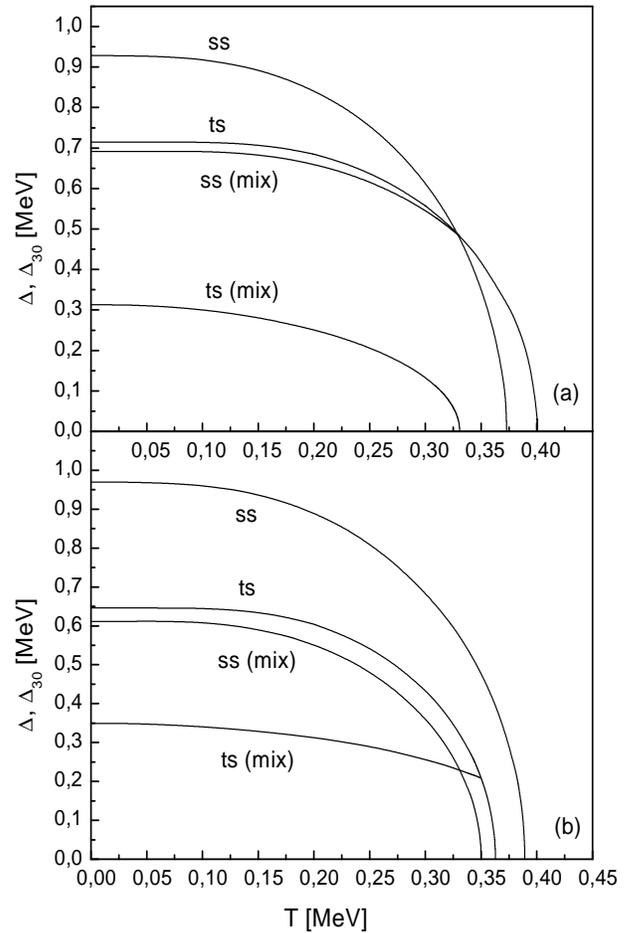}
\caption{Order parameters
$\Delta,\Delta_{30}$ vs. temperature  at density (a)
$\varrho=0.138\ \mbox{fm}^{-3}$ and (b) $\varrho=0.140\
\mbox{fm}^{-3}$. Other notations are the same as in
Fig.~\ref{fig1}. }\label{fig3}
\end{figure}

To determine the temperature behavior of the order parameters, one
should consider Eqs.~\p{12.1}, \p{12.2}. According to our
analysis we can consider two possibilities, when branching occurs
at density $\varrho$ such that (1) $\varrho_1<\varrho<\varrho_m$
and (2) $\varrho_m<\varrho<\varrho_2$. The results of numerical
calculations are shown in Fig.~\ref{fig3}.
In the first case (Fig.~\ref{fig3}(a)) we have
$\Delta_{30}(T_c)=0, \Delta(T_c)=\Delta^{ss}(T_c)$, and, as in the
case of zero temperature, the one--gap order parameters in the SS
and TS pairing channels are also equal,
$\Delta^{ss}(T_c)=\Delta^{ts}_{30}(T_c)$. In the second case
(Fig.~\ref{fig3}(b))
$\Delta(T_c)=0,\Delta_{30}(T_c)=\Delta^{ts}_{30}(T_c)$. Thus, the
temperature region $T<T_c$ corresponds to anisotropic multi--gap
superfluidity when, together with one--gap solutions, we have
two--gap solutions with nonzero SS and TS order parameters in both
pairing channels.

\section{Thermodynamic Stability}
\begin{figure}[tbp]
\includegraphics[height=7.0cm,width=8.6cm,trim=48mm 148mm 57mm 64mm,
draft=false,clip]{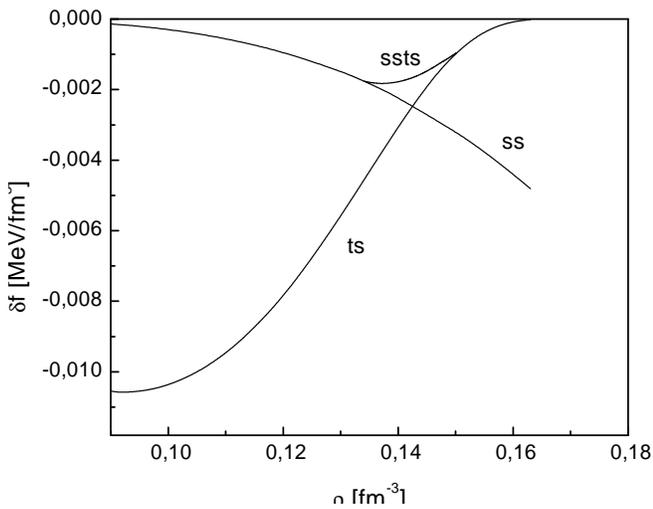} \caption{Free energy density, measured
from that of the normal state, for SS, TS and SS--TS superfluid
states at zero temperature.}\label{fig4}
\end{figure}

\begin{figure}[htbp]
\includegraphics[height=12.6cm,width=8.6cm,trim=47mm 104mm 53mm 50mm,
draft=false,clip]{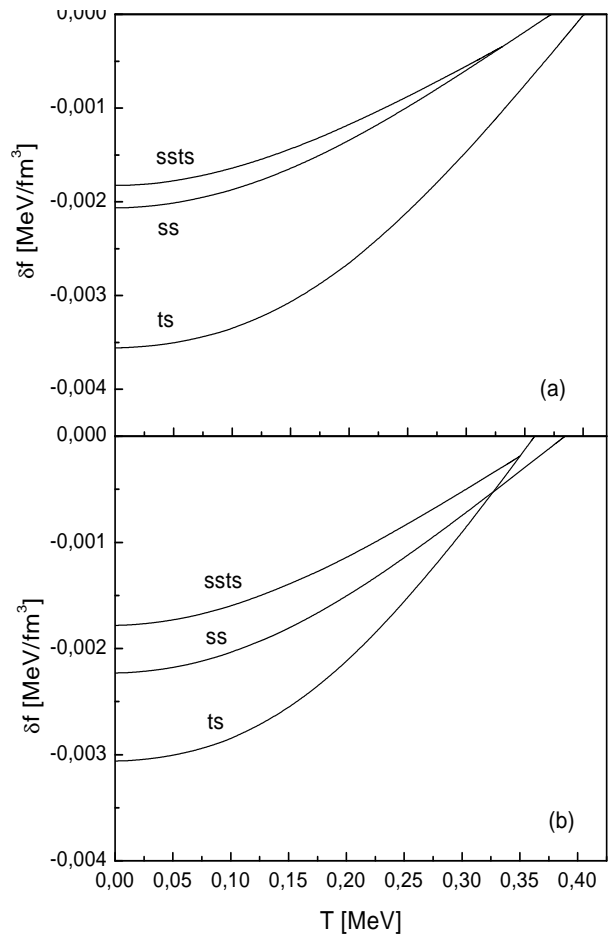} \caption{Difference between free energy
densities of superfluid and normal states vs. temperature at
density (a) $\varrho=0.138\ \mbox{fm}^{-3}$ and (b)
$\varrho=0.140\ \mbox{fm}^{-3}$.  }\label{fig5ab}
\end{figure}

Since we have a few solutions of the self--consistent equations it is
necessary to check which solution is thermodynamically favorable.
For this purpose it is necessary to determine the free energy of the
corresponding states. It consists of two terms,
$F=E(f,g)-TS(f,g)$, where $S$ is entropy of the system. Taking
into account \p{13}--\p{12}, the energy functional \p{14}
 is
\bal E(f,g)
&=2\sum_{\bf{p}}\varepsilon({\bf{p}})(1-\frac{\xi}{2E_+}\tanh{\frac{E_+}{2T}}-
\frac{\xi}{2E_-}\tanh{\frac{E_-}{2T}})\label{29}\\
\quad&-\frac{1}{2}\sum_{\bf{p}}\Bigl\{\frac{|\Delta_+|^2}{E_+}
\tanh{\frac{E_+}{2T}}+
\frac{|\Delta_-|^2}{E_-}\tanh{\frac{E_-}{2T}}\Bigr\},\nonumber\\
\varepsilon({\bf p})&=\frac{{\bf p}^2}{2m},\;
E_\pm=\sqrt{\xi^2+|\Delta_\pm|^2},\;\Delta_\pm=\Delta_{00} \pm
\Delta_{30}.\nonumber\end{align}
The entropy of the system, given in a general theory of superfluid FL
by the expression $S=-\mbox{Tr}\hat f\ln\hat f$ \cite{AKP}, can be
represented in the form \beqstar
S&=-2\sum_{\bf{p}}\Bigl(n_+\ln\,n_++(1-n_+)\ln(1-n_+)\\
&\quad +n_-\ln\,n_-+(1-n_-)\ln(1-n_-)\Bigr),
\eeqstar
 where $n_\pm=\{\exp(E_\pm/T)+1\}^{-1}$. The results of a numerical
 calculation of the free
energy density, measured from that of the normal state, for the
case of  zero temperature, are given in Fig.~\ref{fig4}.
One can see
that in the density interval $\varrho_1<\varrho<\varrho_m$ the TS
superfluid phase is thermodynamically most preferable as
compared with other phases, and at $\varrho_m<\varrho<\varrho_2$
the SS superfluid state wins competition for thermodynamic stability.
In both cases the mixed SS--TS state appears as a result of a
phase transition in density from a thermodynamically less favorable
one--gap superfluid state (SS state, if
$\varrho_1<\varrho<\varrho_m$, and TS state, if
$\varrho_m<\varrho<\varrho_2$) and corresponds to a metastable
state in superfluid nuclear matter. In Fig.~\ref{fig5ab} we
show the difference between the free energy densities of superfluid and
normal states as a function of temperature. Fig.~\ref{fig5ab}(a)
corresponds to the branching of the SS--TS mixed state from the one--gap
SS solution at fixed density in the range
$\varrho_1<\varrho<\varrho_m$, and Fig.~\ref{fig5ab}(b) depicts
branching from the one--gap TS solution at a density in the interval
$\varrho_m<\varrho<\varrho_2$.
As
seen, for all temperatures  $T<T_c$, the  SS--TS superfluid phase
corresponds to a metastable state in superfluid nuclear matter.

{\section*{CONCLUSION} We have considered the possibility of the
formation of an anisotropic multi--gap condensate in superfluid
symmetric nuclear matter, corresponding to the superposition of
states with SS and TS pairing of nucleons. In the SS channel,
pairing occurs with nonzero orbital angular momentum and hence the
energy gap is an anisotropic function of momentum. The
self-consistent equations for such two--gap states differ
essentially from the equations of BCS theory and contain one-gap
solutions (SS and TS) as particular cases. The analysis of the
self-consistent equations at zero temperature in the logarithmic
approximation shows that anisotropic multi--gap superfluid states
can exist only under quite specific restrictions on the coupling
constants $g_{0}$ and $g_{1}$, describing interaction of nucleons
in the SS and TS pairing channels. Since the constants of the
effective interaction depend on density, there are density
domains, where one-gap or anisotropic two-gap solutions exist.
Calculations with the effective SkP interaction, chosen as the
model of the  NN interaction, indicate that two-gap SS--TS states
can  arise in nuclear matter as a result of a phase transition in
density from a one--gap SS or TS state. In the first case at
critical density $g_1>g_0/3$, in the second, the opposite
inequality is valid. Comparing free energies, branching occurs
from thermodynamically less favorable one--gap solution and hence
the anisotropic two--gap superfluid state corresponds to a
metastable state in nuclear matter. Determination of the critical
temperature $T_c$ of the transition to the SS--TS state as a
function of density shows that the corresponding curve consists of
two branches. One of them is related to the appearance of a SS--TS
anisotropic state as a result of branching at $T_c$ from the
one--gap SS solution, another is related to branching from the
one--gap TS solution. Studying the temperature behavior of the
order parameters shows that mixed two--gap solutions exist for
temperatures $T<T_c$. Comparison of free energies leads to the
conclusion that the anisotropic SS--TS phase represents a
metastable state for the whole temperature interval  $T<T_c$.
Calculations show that mixed SS--TS states exist in a density
domain, that is close to nuclear matter saturation density, in
contrast to TS--ST mixed states, which only exist in the low
density domain of nuclear matter.}

{\bf Acknowledgement.} A.I. is grateful for the  hospitality and
support of Rostock University, where a significant part of this work
was completed. Discussions with S. Peletminsky and A.
Yatsenko are gratefully appreciated. A.I. acknowledges the
financial support of STCU (grant No. 1480).
  

\begin{thebibliography}{99}
\bibitem{BM} A. Bohr and B.R. Mottelson, \tit{ Nuclear structure}
(Benjamin, New York, 1969), Vol. 1.
\bibitem{RS} P. Ring and P. Schuck, \tit {The Nuclear Many--Body Problem}
(Springer, New York, 1980).
\bibitem{G} A. Goodman,   Nucl. Phys.   {\bf A352}, 30 (1981); {\bf A352}, 45 (1981);
{\bf A369}, 365 (1981).
\bibitem{RSSL} G. R\"opke, A. Schnell, P. Schuck, and U. Lombardo,  Phys.  Rev.  C {\bf 61},
 024306 (2000).
\bibitem{AFRS}Th.  Alm, B.L.  Friman,
G.  R\"opke, and H. Schulz, Nucl.  Phys.  {\bf A551}, 45 (1993).
\bibitem{BLS} M.  Baldo, U.  Lombardo, and P.  Schuck, Phys.  Rev. C  {\bf
52}, 975 (1995).
\bibitem{ARS}Th.  Alm,
G.  R\"opke, A. Sedrakian, and H. Schulz, Nucl.  Phys.  {\bf
A604}, 491 (1996).
\bibitem{SAL} A.  Sedrakian, T.  Alm, and U.  Lombardo, Phys.  Rev. C
{\bf 55}, R582 (1997).
\bibitem{G1} A. Goodman,   Phys. Rev. C   \tbf{60}, 014311 (1999).
\bibitem{SW} W. Satula and R. Wyss,   Phys.  Rev. Lett. {\bf 86}, 4488
(2001); {\bf 87}, 052504 (2001).
 \bibitem{AIP2} A.I.  Akhiezer, A.A.
Isayev, S.V.  Peletminsky, and A.A.  Yatsenko, Phys.  Lett. B {\bf
451}, 430 (1999).
\bibitem{AAI}  A.A.  Isayev,   Phys.  Rev.  C {\bf 65}, 031302(R) (2002).
       \bibitem{AKP} A.I.
Akhiezer, V.V.  Krasil'nikov, S.V.  Peletminsky, and A.A.
Yatsenko, Phys. Rep.  {\bf 245}, 1 (1994).
       \bibitem{DFT} J.  Dobaczewski, H.  Flocard, and J.
Treiner, Nucl.  Phys.   {\bf A422}, 103 (1984).
\bibitem{RDN} P.-G.  Reinhard, D.J. Dean, W.  Nazarewicz, J. Dobaczewski,
J.A. Maruhn, and M.R. Strayer, Phys.  Rev.  C {\bf 60}, 014316
(1999).
  \bibitem{SYK} R.K.  Su, S.D.  Yang, and T.T.S.  Kuo, Phys.  Rev.  C
{\bf 35}, 1539 (1987).
\bibitem{JK} M.F.  Jiang and T.T.S.  Kuo, Nucl.
Phys.   {\bf A481}, 294 (1988).
\bibitem{AIP} A.I.  Akhiezer, A.A.  Isayev, S.V.  Peletminsky, A.P.
Rekalo, and A.A. Yatsenko, Zh.  Eksp. Teor.  Fiz.  {\bf 112}, 3
(1997) [Sov. Phys.  JETP {\bf 85}, 1 (1997)].
\bibitem{AIPY} A.I.  Akhiezer, A.A.  Isayev, S.V.  Peletminsky,
  and A.A. Yatsenko,   Phys.  Rev.  C {\bf 63}, 021304(R) (2001).
 \end{thebibliography}
   \end{document}